\documentclass[conference]{IEEEtran}
\IEEEoverridecommandlockouts
\usepackage{cite}
\usepackage{amsmath,amssymb,amsfonts}
\usepackage{algorithmic}
\usepackage{graphicx}
\usepackage{textcomp}
\usepackage{xcolor}
\usepackage{mathtools}
\usepackage{subcaption}
\usepackage{xspace}

\definecolor{todocolour}{RGB}{200,45,45} \definecolor{responsecolour}{RGB}{0,0,0} \definecolor{donecolour}{RGB}{30,135,70} 

\newcommand{\nt}[1]{\textcolor{black}{#1}}

\newcommand{\CNOT}{\ensuremath{\text{CNOT}}\xspace}
\newcommand{\CNOTs}{\ensuremath{\text{CNOTs}}\xspace}
\newcommand{\C}{\ensuremath{\mathcal{C}}\xspace}

\usepackage{booktabs} 

\def\BibTeX{{\rm B\kern-.05em{\sc i\kern-.025em b}\kern-.08em
    T\kern-.1667em\lower.7ex\hbox{E}\kern-.125emX}}
\begin{document}

\title{AlphaClifford: Efficient Clifford Synthesis and Transpilation with Model-based RL}
\author{

\IEEEauthorblockN{Daniele {Lizzio Bosco}}
\IEEEauthorblockA{\textit{University of Udine}\\
Udine, Italy \\
 \textit{University  of Naples ``Federico II''}\\
Naples, Italy\\
lizziobosco.daniele@spes.uniud.it}
\and
\IEEEauthorblockN{Jacopo Cossio}
\IEEEauthorblockA{\textit{University of Udine}\\
Udine, Italy \\
cossio.jacopo@spes.uniud.it}
\and
\IEEEauthorblockN{Carla Piazza}
\IEEEauthorblockA{\textit{University of Udine}\\
Udine, Italy \\
carla.piazza@uniud.it}
\and
\IEEEauthorblockN{Giuseppe Serra}
\IEEEauthorblockA{\textit{University of Udine}\\
Udine, Italy \\
giuseppe.serra@uniud.it}
}

\maketitle

\begingroup
\renewcommand{\thefootnote}{}
\footnotetext{© 2026 IEEE. Personal use of this material is permitted. Permission from IEEE must be obtained for all other uses, in any current or future media, including reprinting/republishing this material for advertising or promotional purposes, creating new collective works, for resale or redistribution to servers or lists, or reuse of any copyrighted component of this work in other works.}
\endgroup

\begin{abstract}
    Clifford circuits play a foundational role in quantum computing, particularly due to their importance in quantum error correction and fault-tolerant logical synthesis. While these circuits can be efficiently simulated and represented as symplectic matrices, standard synthesis methods—such as the Aaronson-Gottesman algorithm—often yield sub-optimal circuits with excessively high gate counts. In this work, we introduce AlphaClifford, a model-based Reinforcement Learning framework powered by Monte Carlo Tree Search, designed to efficiently synthesize Clifford circuits from the fundamental gate set composed of H, S, and CNOT. By modeling the state space through the algebraic properties of the symplectic group, AlphaClifford effectively explores this combinatorial space to minimize overall circuit cost. For unconstrained Clifford optimization, our approach achieves a consistent reduction in both total and two-qubit (CNOT) gate counts compared to state-of-the-art synthesis heuristics, despite operating with a strictly less expressive gate set. Furthermore, we demonstrate the broad applicability of our framework on two additional tasks: hardware-constrained Clifford transpilation, where we outperform existing RL-based compilers, and as a post-synthesis optimization component within a full Clifford+T logical synthesis pipeline. Our results underscore that model-based RL is highly effective at addressing the combinatorial complexities of quantum compilation, offering a scalable pathway to mitigate hardware constraints in both near-term and future fault-tolerant quantum devices.
\end{abstract}


\begin{figure*}[!ht]
    \centering
\includegraphics[width=0.99\textwidth,
        trim={0 50 0 260}, clip]{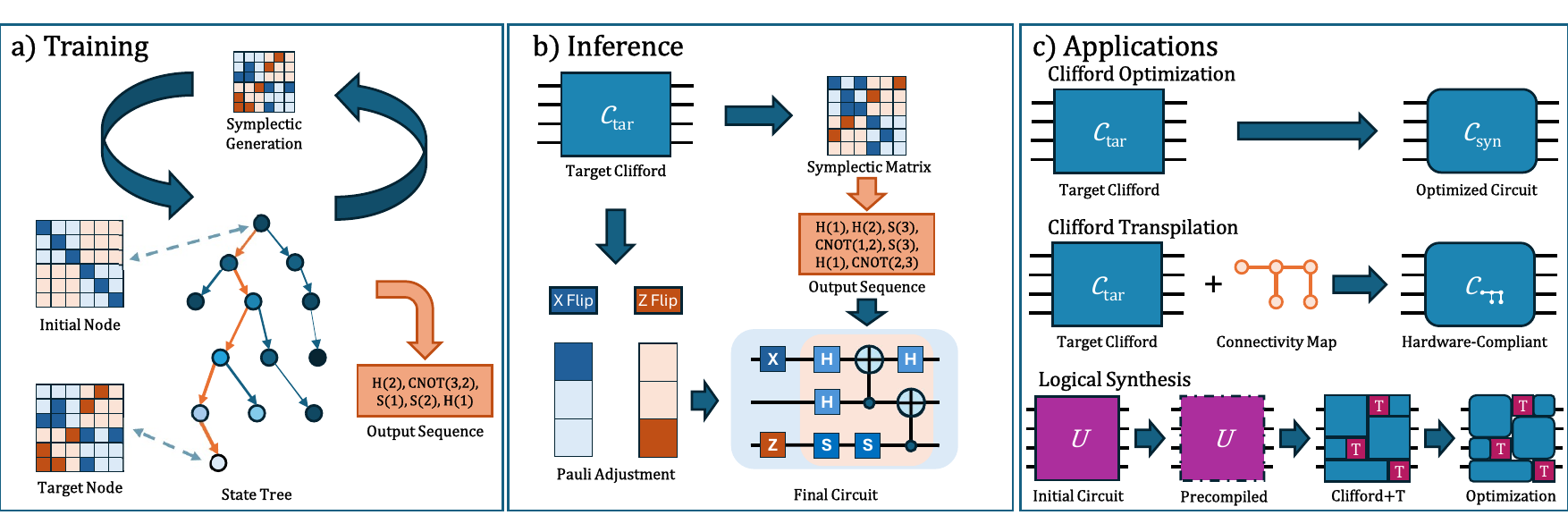}
    \caption{\textbf{a)} We train our model to reproduce a target symplectic matrix. Each symplectic matrix is modelled as the node of a tree, where nodes are connected if the corresponding matrices can be reached with an operation from H, S, and CNOT. We train a MCTS-based model as a pair of value network (for nodes, e.g. lighter blue corresponds to a higher value) and policy network (for edges, e.g. wider arrows corresponds to higher probabilities).  \textbf{b)} Given a target Clifford circuit $C_\text{tar}$, we utilize our model to synthesize a circuit $C$ with the same symplectic matrix, i.e. $P\cdot C\equiv C_\text{tar}$ for a suitable Pauli $P$.
    We then compute the $P$ from initial circuit, and we apply the corresponding gates to the obtained circuit.
    \textbf{c)} AlphaClifford can be applied to a variety of tasks, including  general Clifford circuits optimization, Clifford transpilation to ensure compatibility with a given hardware connectivity map, and as a component in a pipeline for logical synthesis, from a general initial circuit, to the final approximated version in Clifford+T.}
    \label{fig:framework}
\end{figure*}

\section{Introduction}
The field of quantum computing is rapidly approaching the era of ``quantum utility" \cite{Quantum_utility}, marking a pivotal transition where quantum devices cease to be experimental research instruments and begin to execute practical, large-scale applications. 
These applications are expected to span both scientific domains, such as Quantum Chemistry \cite{QC_for_QChem} and High Energy Physics \cite{QC_for_HEP}, and industrial use cases, including discrete optimization \cite{QC_for_OPT} and finance \cite{QC_for_Finance}.

Despite significant recent advancements in quantum hardware capabilities—most notably in error rate reduction and coherence time improvements \cite{google_quantum_ai_suppressing_2023, place_new_2021}—the execution of quantum algorithms remains fundamentally constrained by circuit depth and the total number of physical gates.
To mitigate these hardware limitations, quantum circuit optimization plays a critical role. As equivalent unitary transformations can be realized by different sequences of quantum gates, the objective of quantum circuit synthesis \cite{kitaev_quantum_1997, amy_meet_middle_2013, kliuchnikov_fast_2013, davis_towards_2020, he_unitary_2023} is to find a sequence of gates that implements a target circuit (or its mathematical representation) while minimizing a specific cost metric, such as overall circuit depth or the number of two-qubit entangling gates.

A closely related challenge is quantum circuit transpilation \cite{compilerdesign, compilersmappingrouting}, which involves mapping a logical circuit onto a specific physical architecture. This requires decomposing the target circuit into a native gate set and respecting strict hardware constraints, such as the limited connectivity map of the physical qubits \cite{preskill2018quantumcomputingin}.

The problems of quantum circuit synthesis and transpilation have been extensively explored in the literature, utilizing a wide array of heuristic, algebraic, and search-based techniques \cite{amy_meet_middle_2013, Q2019, Q20192, Q2023}.
Recently, Reinforcement Learning (RL) has emerged as a highly effective paradigm for tackling these computationally hard optimization problems \cite{alphatensor_quantum, wang2024quantumcompilingreinforcementlearning, kremer2025optimizingnoncliffordcountunitarysynthesis, puviani25, QCGame, Kundu_2025, romanello2025CNOT, KremerIBM, cossio2026alphacnotlearningcnotminimization, equivariant_clifford}. By learning complex patterns rather than relying  on heuristic rules, RL-based approaches have demonstrated the ability to discover novel, non-intuitive gate sequences that rule-based compilers might otherwise miss.

In this work, we propose a RL-based technique to address both the synthesis and transpilation tasks specifically for the set of Clifford circuits. Generated strictly by the Hadamard ($H$), Phase ($S$), and Controlled-NOT (\CNOT) gates, Clifford circuits are one of the most studied gate sets, and play a central role in the logical synthesis pipeline, particularly within the context of Clifford+T decomposition \cite{kremer2025optimizingnoncliffordcountunitarysynthesis}.

A key advantage of Clifford gates is that they admit an efficient classical representation via symplectic matrices \cite{AG_for_clifford}. We exploit this compact representation to develop a RL framework designed to learn highly efficient syntheses and transpilation strategies for Clifford circuits.

Inspired by recent successes on similar tasks from AlphaTensor  \cite{alphatensor_quantum} and AlphaCNOT \cite{cossio2026alphacnotlearningcnotminimization}, our method formulates the circuit compilation problem as a tree-based search over the space of symplectic matrices, with the aim of finding the shortest path from an initial matrix to the identity, allowing for a \nt{compact} reconstruction of the corresponding Clifford circuit from the set gate $\{H, S, \CNOT\}$.

To demonstrate the broad applicability of our proposed method, we evaluate it on three different tasks: unconstrained Clifford optimization, Clifford transpilation to a target hardware connectivity map, and as a refinement step in a full logical synthesis pipeline.

We show that our model achieves comparable or higher performance with respect to a variety of different techniques, including up to $56\%$ total gate reduction compared to the Aaronson-Gottesman algorithm \cite{AG_for_clifford} for the general Clifford optimization task, and up to $20\%$ compared to the RL-based Clifford transpiler from \cite{KremerIBM} on the task of transpilation. Our results suggest that AlphaClifford can be applied efficiently in different settings.

The paper is organized as follows. In Section \ref{sec:background} we provide a brief introduction to the symplectic representation of Clifford circuits. In Section \ref{sec:RL} we discuss related works. In Section \ref{sec:methpd} we present our method. Experimental design and results are discussed in Section \ref{sec:exp}. Finally, in Section \ref{sec:conc} we provide some final considerations and suggest possible directions for further works.

\section{Symplectic Representation of Clifford Circuits}
\label{sec:background}
\subsection{Stabilizer Formalism and Tableau Representation}
The $n$-qubit Pauli group $\mathcal{P}_n$ consists of all $n$-fold tensor products of the Pauli matrices $\{I, X, Y, Z\}$ together with the multiplicative phase factors $\{\pm 1, \pm i\}$. The Clifford group $\mathcal{C}_n$ is defined as the normalizer of the Pauli group within the unitary group $U(2^n)$. Specifically, for any Clifford operator $C \in \mathcal{C}_n$ and any Pauli operator $P \in \mathcal{P}_n$, the conjugation $CPC^\dagger$ yields another operator in $\mathcal{P}_n$. The group $\mathcal{C}_n$ is finitely generated by the Hadamard ($H$), Phase ($S$), and Controlled-NOT (\CNOT) gates.

By the Gottesman-Knill theorem \cite{Gottesman1997zz}, Clifford circuits can be efficiently simulated on classical hardware. The standard data structure for this is the tableau representation introduced by Aaronson and Gottesman \cite{AG_for_clifford}. 
The state of an $n$-qubit Clifford circuit is completely characterized by tracing the evolution of $n$ destabilizer generators and $n$ stabilizer generators. In this formalism, the circuit is represented as a $2n \times (2n+1)$ binary matrix $\mathcal{T}$ over the Galois field $\mathbb{F}_2$:
\begin{equation}
    \mathcal{T} = \left[ \begin{array}{cc|c}
    \mathcal{X} & \mathcal{Z} & \mathbf{r}
    \end{array} \right],
\end{equation}
where $\mathcal{X}$ and $\mathcal{Z}$ are $2n \times n$ Boolean matrices denoting the presence of $X$ and $Z$ Pauli operators, respectively. The $2n$-dimensional column vector $\mathbf{r}$ contains Boolean phase bits indicating whether the overall sign of each respective generator is $+1$ ($r_i=0$) or $-1$ ($r_i=1$).

As a simple example, consider a single-qubit system. The initial tableau corresponding to the identity circuit is
\begin{align*}
\mathcal{T}_I = 
\left[
\begin{array}{cc|c}
1 & 0 & 0 \\
0 & 1 & 0
\end{array}
\right],
\end{align*}
where the first row represents the destabilizer generator $X$ and the second row represents the stabilizer generator $Z$.

Applying a Hadamard gate $H$ exchanges $X$ and $Z$, resulting in the tableau
\begin{align*}
\mathcal{T}_H = 
\left[
\begin{array}{cc|c}
0 & 1 & 0 \\
1 & 0 & 0
\end{array}
\right].
\end{align*}
This illustrates how Clifford operations act linearly on the $(\mathcal{X}, \mathcal{Z})$ components of the tableau by transforming Pauli generators.
If we further apply an $X$ gate, the Pauli structure remains unchanged, but the phase of the $Z$ generator is flipped:
\begin{align*}
\mathcal{T}_{XH} = 
\left[
\begin{array}{cc|c}
0 & 1 & 0 \\
1 & 0 & 1
\end{array}
\right].
\end{align*}
Thus, while the matrices $\mathcal{X}$ and $\mathcal{Z}$ encode the Pauli action of the circuit, the vector $\mathbf{r}$ captures the accumulated phase information.

\subsection{Symplectic Matrix Representation}
While the phase vector $\mathbf{r}$ is required to track the exact quantum state (including Pauli phases), the fundamental entangling structure and basis transformations induced by a Clifford circuit are entirely captured by the $2n \times 2n$ binary matrix $M = [\mathcal{X} \;\; \mathcal{Z}]$.

To preserve the commutation relations of the Pauli group, the matrix $M$ must satisfy the symplectic condition over $\mathbb{F}_2$:
\begin{equation}
    M \Omega M^T = \Omega \pmod 2,
\end{equation}
where $\Omega$ is the $2n \times 2n$ block matrix defined as:
\begin{equation}
    \Omega = \begin{bmatrix}
    0 & I_n \\
    I_n & 0
    \end{bmatrix},
\end{equation}
and $I_n$ is the $n \times n$ identity matrix. Consequently, $M$ is an element of the symplectic group $Sp(2n, \mathbb{F}_2)$.

Crucially, there exists a surjective group homomorphism $\phi : \mathcal{C}_n \rightarrow Sp(2n, \mathbb{F}_2)$ that maps a Clifford unitary to its corresponding symplectic matrix. The kernel of this homomorphism is precisely the Pauli group $\mathcal{P}_n$ (ignoring global phases). 

\subsection{Implications for Circuit Synthesis}

In our context of synthesizing Clifford circuits, the homomorphism $\phi$ provides a powerful mathematical abstraction. If an RL agent restricts its objective to finding a sequence of $\{H, S, \text{CNOT}\}$ gates that yields a symplectic matrix $M_{\text{syn}}$ matching a target symplectic matrix $M_{\text{target}}$, the synthesized circuit $C_{\text{syn}}$ and the target circuit $C_{\text{target}}$ are guaranteed to belong to the same coset of the Pauli group.
Mathematically, this establishes an equivalence up to a Pauli layer:
\begin{equation}
    C_{\text{target}} \equiv P \cdot C_{\text{syn}},
\end{equation}
where $P \in \mathcal{P}_n$. 
By tracking the evolution of the phase vector $\mathbf{r}$ alongside the synthesis process, the residual Pauli correction $P$ can be computed efficiently in $\mathcal{O}(n^2)$ time. Therefore, mapping the RL state space to $Sp(2n, \mathbb{F}_2)$ rather than the full Clifford group $\mathcal{C}_n$ reduces the search space by a factor of $2^{2n}$, simplifying the learning process.

\section{Related Work}
\label{sec:RL}

The problem of quantum circuit synthesis and compilation is extremely important for mitigating  hardware limitations of current devices, and the literature addressing this challenge is extensive \cite{yan2024quantum}.

Within the context of Clifford circuits, the historical baseline is the well-known Aaronson-Gottesman algorithm \cite{AG_for_clifford}, which demonstrated that this class of circuits can be efficiently simulated and synthesized in polynomial time with gates from $\{H, S, \text{CNOT}\}$.

However, standard approaches usually generate sub-optimal circuits with respect to the number of gates. To address this, various advanced heuristic methods have been proposed.

For instance, the approaches described in \cite{webster} (such as A*, greedy algorithms, and Volanto \cite{volanto}) leverage a richer generating set to \nt{reduce the gate count in the circuit.} Another approach is the one proposed in \cite{Bravyi2021cliffordcircuit}, where Clifford-specific template matching is combined with symbolic peephole optimization, obtaining relevant reductions in CNOT count.

It is also worth noting that the literature offers solutions for the optimal and exact syntheses of Clifford circuits \cite{bravyi2022optimalclifford, shaik2025cnotoptimal}. However, finding the absolute minimum sequence of gates generally requires exponential time, drastically limiting the scalability of these methods to a very small number of qubits and rendering them unsuitable for practical, large-scale applications.

Recently, many  Reinforcement Learning (RL) based solutions have been proposed to address tasks related to circuit synthesis, optimization, and compilation, including state preparation \cite{Kundu_2025}, quantum architecture search \cite{kundu_tensorrl_qas},  and “pre-compilation” in logical synthesis pipelines in \textsc{Q-PreSyn} \cite{qpresyn}.

The synthesis of Clifford circuits is particularly well-suited for RL applications: the algebraic properties of these circuits provide a compact and highly efficient state representation, usually as a matrix of polynomial size.
Initially, many works were based on \textbf{model-free} algorithms, such as  Proximal Policy Optimization \cite{schulman2017ppo} (PPO), which learn directly by interacting with the environment without building an internal model of the dynamics. 

 Notable among these is the recent work \cite{KremerIBM}, which leverages RL for the routing and synthesis of Clifford circuits, yielding transpilers capable of significantly outperforming traditional heuristics like SABRE \cite{zou_lightsabre_2024}. 
 Another  impactful model-free application is provided in \cite{puviani25} in the context of quantum circuit discovery for fault-tolerant logical state preparation, where RL was used to discover entirely new protocols that surpass the efficiency of manually designed circuits.
Additionally, in \cite{romanello2025CNOT} the authors propose a PPO-based approach to optimize CNOT count in linear reversible circuits.
 Concurrent work \cite{equivariant_clifford} also addresses Clifford circuit synthesis using model-free RL. Their approach formulates synthesis as the reduction of a symplectic tableau to the identity using PPO, and introduces a permutation-equivariant, size-agnostic neural architecture that generalizes across different numbers of qubits, allowing to learn a scalable policy for unconstrained Clifford synthesis.

Despite the significant successes of model-free methods, recent developments demonstrate that when the mathematical structure of the problem allows it, \textit{model-based} algorithms, usually based on Monte Carlo Tree Search \cite{coulom2007} (MCTS), offer vastly superior capabilities for exploring complex combinatorial spaces.

 In this regard, AlphaTensor-Quantum \cite{alphatensor_quantum} paved the way by demonstrating how the CNOT+$T$ component of a circuit can be optimized using search techniques originally designed for board games \cite{alphazero}. Similarly, AlphaCNOT \cite{cossio2026alphacnotlearningcnotminimization} provided a model to address the problems of CNOT optimization in both unconstrained and constrained (i.e., depending on a connectivity map) settings.

 Finally, another relevant contribution  is \textit{QuSynth} \cite{doherty_fast_2026}, in which the task of stabilizer state preparation is  reformulated  by using graph-state representations, and addressed by combining  RL with MCTS for structured look-ahead.

Our work, \textbf{AlphaClifford}, builds on these developments by bringing the strengths of search-informed reinforcement learning to the synthesis of general Clifford circuits. In particular, we draw inspiration from the AlphaTensor and AlphaCNOT frameworks, adapting their core ideas to exploit the algebraic structure of Clifford operations and further advance the state of the art in this domain.

\section{AlphaClifford}
\label{sec:methpd}
A popular class of Reinforcement Learning (RL) approaches is classified as \emph{model-free}, i.e., they learn directly from interactions with the environment and do not leverage an explicit model of its dynamics. Model-free approaches, such as  Proximal Policy Optimization \cite{schulman2017ppo} (PPO), rely on sequences of state–action–reward transitions to iteratively update the policy via gradient-based optimization. 

Conversely, \emph{model-based} RL algorithms construct a representation of the environment to simulate future outcomes \cite{kocsis2006MCTS}. Within complex combinatorial spaces, such as the set of symplectic matrices, their capacity for structured and efficient exploration enables them to obtain improved performance over both classical heuristics and model-free algorithms.

Following the latter principle, and inspired by recent applications such as AlphaTensor \cite{alphatensor_quantum} and AlphaCNOT \cite{cossio2026alphacnotlearningcnotminimization}, we propose \textbf{AlphaClifford}, a novel model-based approach for exploring the combinatorial space of  symplectic matrices associated with Clifford circuits.

Our method, illustrated in Fig. \ref{fig:framework}, is organized into the following steps:
\begin{itemize}
    \item First, we model the space of symplectic matrices as a search tree (Fig. \ref{fig:framework}a), where each node represents a specific symplectic matrix and directed edges correspond to the application of Clifford gates. In this representation, two nodes are connected if the corresponding matrices can be reached through a Clifford gate operation.
    \item Second, we train a model to learn the structure of this space by predicting sequences of gate operations that transform the initial symplectic matrix into the target one. The training procedure, structured after the Monte Carlo Tree Search pattern \cite{coulom2007}, is detailed in Section \ref{SubSec:MCTS}.

    \item Third, once trained, the model is used to synthesize Clifford circuits (Fig. \ref{fig:framework}b) by generating a new sequence of gates from a target circuit. The generated sequence of Clifford gates will have the same symplectic matrix as the one associated to the target circuit. 
\end{itemize}

\begin{figure*}[!ht]
    \centering
\includegraphics[width=0.99\textwidth,
        trim={0 62 0 270}, clip]{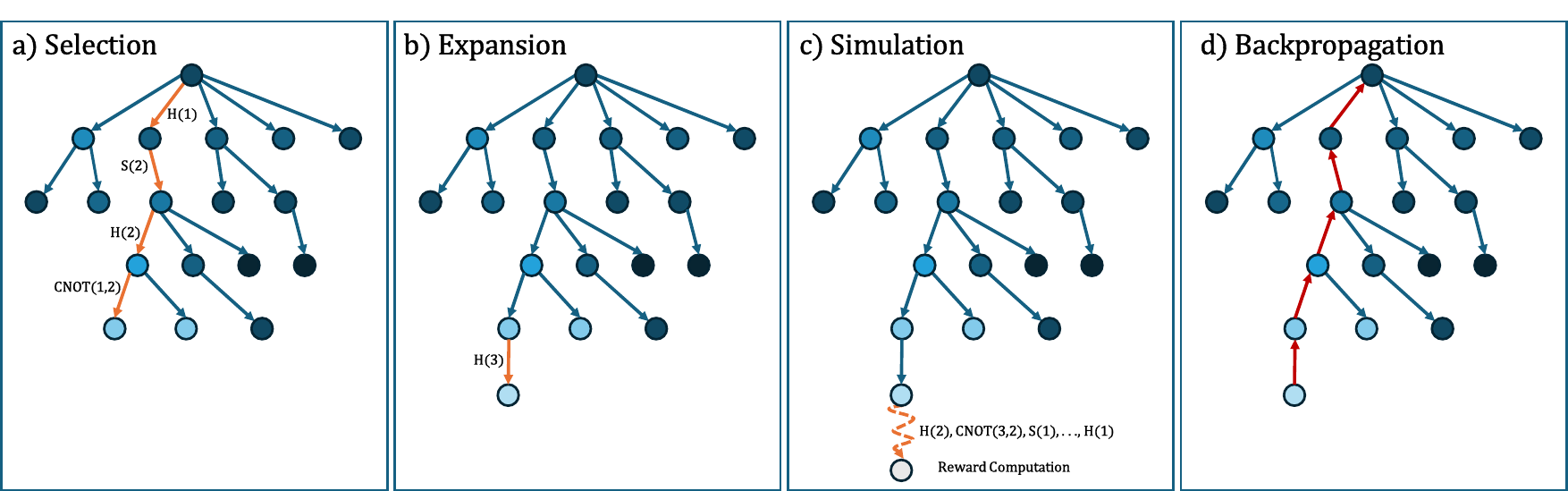}
    \caption{Overview of the Monte Carlo Tree Search (MCTS) phases for Clifford circuit synthesis. The search space is explored by iteratively executing four distinct stages: (a) Selection: The tree is traversed from the root (representing the target symplectic matrix \C) to a leaf node. Actions, corresponding to Clifford gates $H, S,$ and \CNOT, are chosen by balancing exploration and exploitation using the UCT formula. (b) Expansion: When reaching a leaf node, the tree is expanded by the addition of one or more child nodes, each representing the state resulting from a possible Clifford operation. (c) Simulation: A rollout is performed from the newly expanded node \nt{based on the policy network}. A sequence of gates is sampled until a terminal state is reached, at which point \nt{the evaluation of the value network} is computed based on the proximity to the identity matrix. (d) Backpropagation: The reward obtained during simulation is propagated upward to the root, updating the value estimates and visit counts for all nodes along the selected path to refine future selection steps.}
    \label{fig:framework_MCTS}
\end{figure*}

\subsection{Modeling the problem of symplectic matrices}\label{sub:sec:problem_modeling}

The search space of our problem, represented by the set of symplectic matrices, is explored through a tree-based approach. For a given problem dimension $n$ (corresponding to the number of qubits), given a target Clifford circuit $C$ represented by its symplectic matrix $M_{C} \in Sp(2n, \mathbb{F}_2)$, and possibly a connectivity map $\mathcal{M}$, our aim is to determine a sequence of Clifford gates equivalent to $C$. 

To this end, we construct a search tree as follows:
\begin{itemize}
    \item The root node represents the identity matrix $I_{2n}$, corresponding to the empty circuit;
    \item Each node $N$ is labeled by a symplectic matrix $M$ and has one child for each admissible Clifford operation $g \in \{H_i, S_i, \CNOT_{ij}: i,j\in\{1,\dots,n\}\}$   (subject to $\mathcal{M}$). The child node is labeled by the matrix obtained by applying $g$ to $M$, i.e., by the product of $M$ with the matrix corresponding to $g$;
    \item The terminal nodes (i.e., the leaves) are labeled by the target symplectic matrix $M_{C}$.
\end{itemize}

In the unconstrained setting, each node has $\Theta(n^2)$ children due to the possible $\CNOT$ operations between pairs of qubits, along with single-qubit gates. In the case of limited connectivity constraints, only a subset of 2-qubit operations is allowed.

\subsection{Inverting the representation: from the target to the identity}

Training an RL agent with the forward representation described in Section~\ref{sub:sec:problem_modeling} would require providing, at each step \(k\), both the current symplectic matrix \(M_k\) (encoding the partial circuit \(g_1,\dots,g_k\)) and the fixed target matrix \(M_C\). This increases the input dimensionality and, moreover, the target matrix contributes little information during an episode, as it remains constant. To avoid this, we adopt an inverted formulation that embeds the target directly into the state representation.

At step \(k\), the observation is defined as
\[
M_C\, M_k^{-1},
\]
where \(M_k^{-1}\) is the inverse of the current symplectic matrix over \(\mathbb{F}_2\), corresponding to the inverse circuit. Under this representation, synthesizing the target Clifford circuit is equivalent to reaching the identity:
\[
M_C\, M_C^{-1} = I_{2n}.
\]
Hence, each episode begins in the state \(M_C\), and the agent’s task is to reach the identity matrix. This turns the problem into a goal-reaching task with a fixed terminal state and allows us to provide a single observation matrix to the model, rather than a pair \((M_k, M_C)\).

When a gate \(g_t\) is applied in the forward construction of the circuit, the corresponding update in the inverted representation must use the adjoint action. Since
\[
M_t = M_{g_t}\, M_{t-1} \qquad \Longrightarrow \qquad 
M_t^{-1} = M_{t-1}^{-1} M_{g_t^\dagger},
\]
the state transition in the inverted formulation becomes
\[
M_C\, M_{t-1}^{-1} \;\longrightarrow\; M_C \left(M_{t-1}^{-1} M_{g_t^\dagger}\right).
\]
Thus, although the agent selects forward Clifford operations, the evolution of the internal representation proceeds by right-multiplication with the corresponding symplectic matrices, consistently reflecting the reversed dynamics.
\subsection{Monte Carlo Tree Search Framework}\label{SubSec:MCTS}
We adopt a Monte Carlo Tree Search \cite{coulom2007} (MCTS) strategy to efficiently explore the space of symplectic matrices associated with Clifford circuits (see Fig.~\ref{fig:framework_MCTS}). The procedure is structured into four main phases:

\begin{enumerate}
    \item \textbf{Selection:} Starting from the root node corresponding to the target symplectic matrix $M_{C}$, the tree is traversed by iteratively selecting actions (i.e., Clifford gates in $\{H, S, \CNOT\}$) according to a tree policy. In particular, we adopt the Upper Confidence Bound for Trees (UCT) criterion \cite{kocsis2006MCTS}, which selects the action $a$ maximizing
    \begin{equation}
            \mathrm{UCT}(a) = Q(s,a) + c \sqrt{\frac{\log N(s)}{N(s,a)}},
        \end{equation}

    where $Q(s,a)$ is the estimated value of taking action $a$ in state $s$, $N(s)$ is the visit count of node $s$, $N(s,a)$ is the number of times action $a$ has been selected from $s$, and $c$ is an exploration constant. This criterion balances exploration of less-visited actions and exploitation of high-value ones;

    \item \textbf{Expansion:} When a non-terminal node with unexplored actions is reached, the tree is expanded by applying a new admissible Clifford operation, thus generating a new child node. A node is terminal if it corresponds to the identity matrix or if other stopping criteria are met (e.g., maximum depth). In the presence of connectivity constraints, only allowed $\CNOT$ operations are considered;

    \item \textbf{Simulation:} From the expanded node, a rollout is performed by sampling a sequence of Clifford gates according to \nt{the policy network} until a terminal node is reached. The resulting trajectory is then \nt{processed by the value function to estimate the quality of the new node};

    \item \textbf{Backpropagation:} The outcome of the simulation is propagated back along the selected path, updating visit counts and value estimates for each node involved in the traversal.
\end{enumerate}

\subsection{Training Details}\label{Subsec:training_details}

\nt{The training consists of constructing a different random target matrix for each episode, and asking the model to ``reach'' the identity matrix starting from the inverse of the target with the fewest amount of moves. Each ``move'', corresponding to a \textit{step} in the environment, represents the application of a feasible gate to the current circuit. Feasible gates are $H$, $S$, and \CNOTs, possibly with constraints based on topology connectivity. An episode finishes after reaching the identity, or when the agent perform $n_{\text{max}}$ moves without reaching the identity.}

We adopt a naive reward function to guide the learning process. At each step a penalty of $-1/\sqrt{n}$ is applied to limit long sequences and a terminal reward of $+n$ is assigned when reaching the identity matrix. This ``non-informed'' strategy is fundamental in our approach: without relying on problem-specific heuristics, the model itself naturally learns to find shorter decompositions. In particular, the training objective implicitly minimizes the total number of applied gates, which also implies a reduction of \CNOTs. Note that different rewards could be used to minimize different target functions: for example, we can select a higher penalty to reduce the CNOT count. 

We employ a linear curriculum learning strategy \cite{bengio2009curriculumlearning} based on the (expected) difficulty of the instances. In practice, we generate each target symplectic matrix by selecting random $k$ Clifford gates, where $k$ is parameter we can tune during training. 

More in detail, we divide our training in tranches. Starting from $k=1$, we train our model on a tranche of difficulty $k$. At the end of the tranche, we increase $k$ if the model reached a solving rate of at least $95\%$, i.e., the model is able to correctly reproduce most matrices. We stop the training after the max difficulty of $k_{\max} = 4n^2$ is reached.

\subsection{Inference}
Once the model has been trained to synthesize symplectic matrices of size $2n\times2n$, optionally subject to hardware connectivity constraints, it can be deployed for inference. Specifically, given a target Clifford circuit $C_\text{tar}$ characterized by a symplectic matrix $M_\text{tar}$, our model generates a sequence of gates forming a synthetic circuit $C_\text{syn}$ that yields the identical symplectic matrix. Consequently, $M_\text{tar} = M_\text{syn}$, which establishes that the target and synthetic circuits are equivalent up to a Pauli operator (i.e., $C_\text{tar} \equiv P C_\text{syn}$ for a pauli $P$).

The complete inference procedure for a given target circuit proceeds as follows: (i) efficiently compute its target symplectic matrix $M_\text{tar}$; (ii) query the trained model to synthesize a gate sequence corresponding to this matrix; (iii) efficiently compute the phase discrepancy—namely, the signs of the stabilizers and destabilizers—between the target and synthetic circuits; and (iv) apply the appropriate $X$ and $Z$ gates to correct the phase, yielding a fully equivalent circuit. A visual overview of this pipeline is provided in Fig. \ref{fig:framework}b.

\subsection{Applications}
\label{applications}

We identify three primary applications where our model can be efficiently deployed: Clifford optimization (Fig. \ref{fig:framework}c, top), Clifford transpilation (Fig. \ref{fig:framework}c, middle), and as a subroutine within a standard logical synthesis pipeline (Fig. \ref{fig:framework}c, bottom). These settings are described below and are experimentally demonstrated in Sections \ref{sub:exp:opt}, \ref{sub:exp:trans}, and \ref{sub:exp:pipe}, respectively.

\subsubsection{Optimization}
A direct application of our framework is Clifford circuit optimization. Given a target Clifford circuit, the objective is to find an equivalent representation that minimizes the total number of two-qubit gates. More broadly, the model can be tailored during training to minimize alternative target metrics, such as overall circuit depth. \nt{Beyond full-circuit synthesis, AlphaClifford can also be employed as a peephole  optimizer \cite{peephole} for Clifford circuits involving a larger number of qubits. Starting, for example, from a circuit synthesized using the Aaronson–Gottesman algorithm, one can extract contiguous subcircuits whose support contains at most $n$ qubits, resynthesize each block using the corresponding n-qubit AlphaClifford model, and replace the original block whenever the selected cost is reduced. By iteratively applying this procedure, models trained on relatively small problem sizes can be used to optimize substantially larger circuits, although the resulting improvements are local and do not imply global optimality.}
\subsubsection{Transpilation}

We further evaluate our approach on the task of quantum circuit transpilation. Given $n$ qubits and a hardware connectivity map—represented as a graph where nodes correspond to physical qubits and edges denote allowable direct interactions—the goal of transpilation is to transform a logical circuit into an equivalent physical circuit that strictly adheres to these connectivity constraints. This is usually achieved by mapping logical qubits to physical qubits and inserting additional gates (e.g., SWAP routing) to ensure all two-qubit operations occur between adjacent nodes. 
Addressing this task is critical for near-term quantum execution, as many hardware platforms (such as superconducting architectures \cite{superconductiv}) only permit multi-qubit operations between physically coupled qubits. 

Similar to circuit optimization, the objective is to find a valid transpilation that minimizes a specific cost function, typically the two-qubit gate count or the overall circuit depth. Furthermore, for realistic hardware deployments, the cost function can be expanded to incorporate device-specific characteristics, such as penalizing the use of qubits or couplers with high error rates. 
To accomplish this within our framework, the model's action space during training is constrained to the specific connectivity map of the target hardware connectivity map.

\subsubsection{Optimizing Logical Synthesis Pipeline}

A crucial task in fault-tolerant quantum computing is logical synthesis, which involves approximating a target unitary using a discrete, universal gate set such as Clifford+T \cite{kitaev_quantum_1997}. Unlike pure Clifford synthesis, optimal exact synthesis over universal gate sets is a notoriously difficult problem, with worst-case computational complexity scaling exponentially with respect to the number of qubits \cite{sk_alg, nielsen2000quantum}. To address this limitation, compilers typically employ a ``peephole'' synthesis strategy, decomposing the target circuit into much smaller, tractable sub-blocks (typically acting on two or three qubits) \cite{Bravyi2021cliffordcircuit}.

A standard logical synthesis pipeline generally operates in successive stages. First, a pre-processing step can applied to the logical circuit. This may involve decomposing large multi-qubit gates into smaller two- or three-qubit instances (e.g., via recursive Cartan decomposition \cite{cartan_sur_1926, wierichs_recursive_2025}) or performing structural pre-synthesis edits to make the circuit more amenable to compilation \cite{qpresyn}. Next, a local synthesis algorithm compiles each of these individual sub-blocks into a Clifford+T representation. 

Our framework can be integrated into the final stage of this pipeline as a post-synthesis optimizer. Once the full circuit has been compiled into Clifford+T, we can extract the largest contiguous blocks of purely Clifford gates. Our model can then be applied directly to these extracted Clifford sub-circuits to find equivalent representations that minimize the total gate count, effectively compressing the circuit without altering the expensive T-gate layout. 

While our experimental evaluation in Section \ref{sub:exp:pipe} demonstrates this capability on the relatively small Clifford blocks generated by standard peephole synthesis, our approach can be applied more in general in any kind of compilation task involving large blocks of consecutive Clifford gates.

\section{Experimental Results}
\label{sec:exp}

In the following, we describe the experimental evaluations and the obtained results for the tasks of Clifford Optimization (Section \ref{sub:exp:opt}), Clifford Transpilation (Section \ref{sub:exp:trans}), and optimization in the logical synthesis pipeline (Section \ref{sub:exp:pipe}).

Both policy and value networks share the same architecture, i.e., a 12-layer fully connected network with 256 hidden units per layer. We structured the net using 4 residual blocks with skip connections every three layers.
We trained a different model for each number of qubits $n$, and for each selected connectivity map  for the transpilation task. The training was executed in tranches of different complexity (computed as the number of generating gates) as described in Section \ref{Subsec:training_details}, 
 and increasing once the solving threshold of 95\% is achieved for the tranche, up to a complexity of $4n^2$.

\subsection{Clifford Optimization}
\label{sub:exp:opt}

As a first task, we evaluate the ability of our model to efficiently synthesize sufficiently complex Clifford circuits. This problem can be reformulated as a Clifford optimization task: we start from a random Clifford circuit generated by \texttt{qiskit} \cite{qiskit}, and we synthesize an equivalent circuit with gates from $\{H, S, \text{CNOT}\}$. 

For $n$ qubits with $n\in\{3,4,5,6,7\}$, we evaluate our model on $100$ Clifford circuits  involving up to $4n^2$ gates. \nt{These experiments evaluate full synthesis, in which the complete target circuit is provided to a single model. Note that, as discussed in Section \ref{applications}, the same models can be applied as local peephole optimizers to blocks of larger Clifford circuits.}

To evaluate our approach, we consider both the average total number of gates, and the number of two-qubit gates (in our case, CNOTs) required to synthesize a circuit. We evaluate our model both one shot, and after $10$ repetitions, i.e. we sample our model $10$ times, and select the circuit with lowest amount of gates. We denote the latter as AlphaClifford$_{10}$. Our results are reported in Table \ref{tab:exp1_combined}.

As comparison, we consider first the well-known Aaronson-Gottesman algorithm \cite{AG_for_clifford}, as implemented in \texttt{qiskit} \cite{qiskit}), and then three different heuristic models ($A^*$, greedy, and Volanto \cite{volanto}), as provided in \cite{webster}. These three methods provide a decomposition in a block of SWAP gates, followed by a block of one and two qubit \textit{transvection} gates \cite{transvection}. Given a $n$-qubit Pauli operator $P$, the corresponding transvection can be defined as \begin{equation}
      T_P\coloneqq \exp\left( \frac{i\pi}{4}(I_n-P)\right).
  \end{equation}

  Note that by allowing a richer gate set provides an unfair comparison versus our approach that is constrained to the less expressive set $\{H, S, \text{CNOT}\}$. Nevertheless, we include this evaluation, where the SWAP layer is first decomposed in three CNOTs.
  For completeness, we also consider the CNOT count obtained by the same three methods when the obtained circuits are decomposed in $\{H, S, \text{CNOT}\}$ by the default transpiler provided by qiskit. \nt{Finally, we report the same metrics obtained by the qiskit \textit{greedy} method, and by the well-known \texttt{stim} software for stabilizer simulation \cite{stim}.}
\begin{table}[htb]
    \centering
    \begin{tabular}{lccccc}
        \toprule
        & \multicolumn{5}{c}{Problem Size ($n$)} \\
        \cmidrule(lr){2-6}
        Approach & 3 & 4 & 5 & 6 & 7 \\
        \midrule
        \multicolumn{6}{c}{\textbf{Total Gates}} \\
        \midrule
        Aaronson-Gottesman & 19.99 & 25.89 & 38.05 & 51.37 & 65.61 \\
        \textcolor{responsecolour}{Qiskit greedy} & \textcolor{responsecolour}{13.57} & \textcolor{responsecolour}{23.03} & \textcolor{responsecolour}{34.75} & \textcolor{responsecolour}{47.77} & \textcolor{responsecolour}{62.18}\\
        \textcolor{responsecolour}{Stim} & \textcolor{responsecolour}{21.27} & \textcolor{responsecolour}{36.28} & \textcolor{responsecolour}{54.70} & \textcolor{responsecolour}{72.82} & \textcolor{responsecolour}{96.39}\\
        \addlinespace
        A* (with transv.) & 12.97 & 19.56 & 26.54 & 35.04 & 43.15 \\
        Greedy (with transv.) & 13.03 & 19.40 & 27.42 & 36.08 & 45.61 \\
        Volanto (with transv.) & 12.95 & 19.93 & 28.65 & 38.12 & 48.14 \\
        \addlinespace
        A* ($H, S, \text{CNOT}$) & 19.49 & 33.47 & 49.90 & 68.33 & 89.47 \\
        Greedy ($H, S, \text{CNOT}$) & 19.17 & 33.21 & 51.44 & 72.16 & 98.29 \\
        Volanto ($H, S, \text{CNOT}$) & 20.26 & 37.30 & 59.25 & 87.19 & 117.57 \\
        \addlinespace
        \textbf{AlphaClifford} & 8.85 & 13.90 & 20.66 & 29.56 & 40.05 \\
        \textbf{AlphaClifford$_{10}$} & \textbf{8.78} & \textbf{13.39} & \textbf{19.49} & \textbf{27.11} & \textbf{36.75} \\
        \midrule
        \multicolumn{6}{c}{\textbf{2-Qubit Gates}} \\
        \midrule
        Aaronson-Gottesman & \textbf{3.65} & 10.41 & 16.17 & 22.43 & 28.59 \\
        \textcolor{responsecolour}{Qiskit greedy} & \textcolor{responsecolour}{4.35} & \textcolor{responsecolour}{8.01} & \textcolor{responsecolour}{13.01} & \textcolor{responsecolour}{18.79} & \textcolor{responsecolour}{24.95}\\
        \textcolor{responsecolour}{Stim} & \textcolor{responsecolour}{7.29} & \textcolor{responsecolour}{13.20} & \textcolor{responsecolour}{20.32} & \textcolor{responsecolour}{29.49} & \textcolor{responsecolour}{40.35}\\
        \addlinespace
        A* (with transv.) & 6.34 & 10.63 & 15.72 & 21.39 & 27.32 \\
        Greedy (with transv.) & 6.28 & 10.60 & 16.55 & 22.71 & 29.86 \\
        Volanto (with transv.) & 6.38 & 11.16 & 17.36 & 24.38 & 32.62 \\
        \addlinespace
        A* ($H, S, \text{CNOT}$) & 8.84 & 15.38 & 23.10 & 31.89 & 41.59 \\
        Greedy ($H, S, \text{CNOT}$) & 8.78 & 15.39 & 24.39 & 34.19 & 46.28 \\
        Volanto ($H, S, \text{CNOT}$) & 9.19 & 17.04 & 27.37 & 39.85 & 54.23 \\
        \addlinespace
        \textbf{AlphaClifford} & 4.25 & 7.50 & 11.85 & 17.47 & 25.53 \\
        \textbf{AlphaClifford$_{10}$} & 4.06 & \textbf{6.96} & \textbf{10.98} & \textbf{15.82} & \textbf{23.06} \\
        \bottomrule
    \end{tabular}
    \caption{Average number of total and 2-Qubit gates for different optimization methods over varying problem sizes. Best results for each $n$ are highlighted in bold.}
    \label{tab:exp1_combined}
\end{table}

We observe that our model achieves the lowest amount of total gates, and two qubit gates, in most cases. Note that all methods were able to synthesize correctly each instance, with the exception of Greedy method that was not able to synthesize a few instances of larger dimensions (in this case, the average reported in Table \ref{tab:exp1_combined} is considered only on the solved instances).  

We stress that the advantage compared to the methods in \cite{webster} is obtained with a less expressive gate. We expect our method to obtain even better performance when allowing more expressive gate sets.

\subsection{Transpilation}
\label{sub:exp:trans}

\begin{table*}[htb]
    \centering
    \small 
    \begin{tabular}{l cccccc cccccc}
        \toprule
        & \multicolumn{6}{c}{\textbf{L Maps}} & \multicolumn{6}{c}{\textbf{T Maps}} \\
        \cmidrule(lr){2-7} \cmidrule(lr){8-13}
        
        & \multicolumn{2}{c}{3L} & \multicolumn{2}{c}{4L} & \multicolumn{2}{c}{5L} & \multicolumn{2}{c}{4T} & \multicolumn{2}{c}{5T} & \multicolumn{2}{c}{6T} \\
        \cmidrule(lr){2-3} \cmidrule(lr){4-5} \cmidrule(lr){6-7} \cmidrule(lr){8-9} \cmidrule(lr){10-11} \cmidrule(lr){12-13}
        
        Approach & Tot. & CNOT & Tot. & CNOT & Tot. & CNOT & Tot. & CNOT & Tot. & CNOT & Tot. & CNOT \\
        \midrule
        RL-Qiskit$_{1}$ & 14.52 & 4.91 & 26.80 & 10.51 & 36.95  & 16.78 & 26.38 & 9.60 & 33.63 & 15.15 & 50.85 & 24.41 \\
        RL-Qiskit$_{10}$ & 12.19 &\textbf{4.60} & 22.90 & \textbf{9.54}  & 33.31 & \textbf{16.08} & 21.58 & \textbf{8.43} & 30.92 & \textbf{14.29} & 46.11 & \textbf{22.74} \\
        \addlinespace

        \textbf{AlphaClifford$_{1}$} & 9.83 & 5.15 & 17.49 & 10.66 & 29.07 & 18.45 & 16.16 & 9.51 & 26.27 & 16.82 & 42.35 & 27.49 \\
        \textbf{AlphaClifford$_{10}$} & \textbf{9.81} & 4.99 & \textbf{17.08} & 10.08 & \textbf{27.58} & 17.31 & \textbf{15.76} & 8.96 & \textbf{25.04} & 15.56 & \textbf{38.84} & 25.10 \\
        \bottomrule
    \end{tabular}
    \caption{Performance of RL-Qiskit \cite{qiskit_ai_transpiler_cliffords} and AlphaClifford for the transpilation task. Results report average total (Tot.) and CNOT gate counts for L and T connectivity maps. Best values for metric are reported in bold.}
    \label{tab:LT_topology_combined}
\end{table*}

To evaluate our model, for each number of qubits and connectivity map, we generate a random Clifford circuit without connectivity constraints. We then use a model trained on the specific connectivity map to generate an equivalent circuit by using the gate set $\{H, S, \text{CNOT}\}$, where a CNOT can be selected only between pairs of qubits allowed by the connectivity map.

\begin{figure}[htbp]
    \centering
    \begin{subfigure}[b]{0.32\columnwidth}
        \centering
        \includegraphics[width=\textwidth]{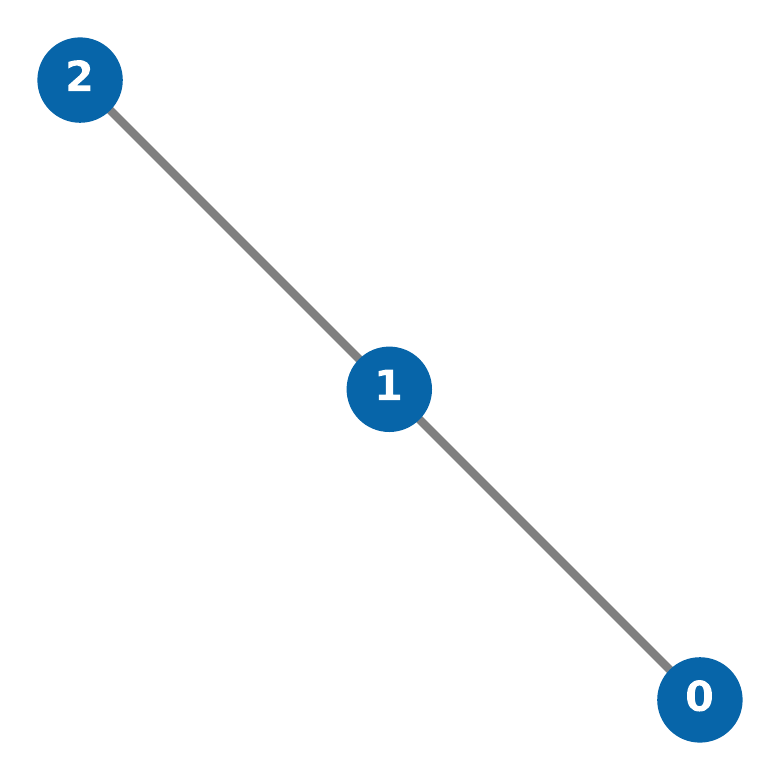}
        \caption{3L}
    \end{subfigure}\hfill
    \begin{subfigure}[b]{0.32\columnwidth}
        \centering
        \includegraphics[width=\textwidth]{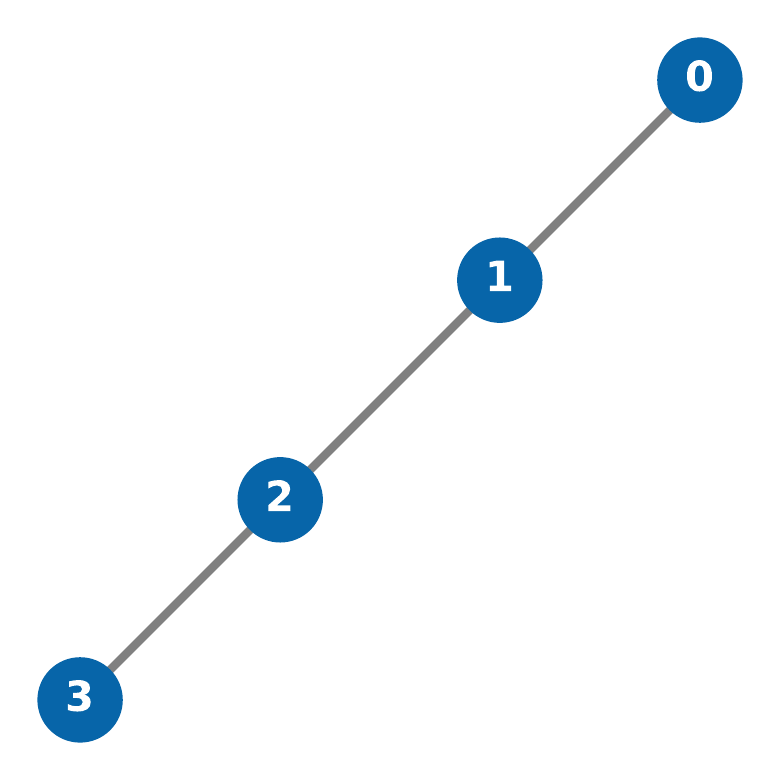}
        \caption{4L}
    \end{subfigure}\hfill
    \begin{subfigure}[b]{0.32\columnwidth}
        \centering
        \includegraphics[width=\textwidth]{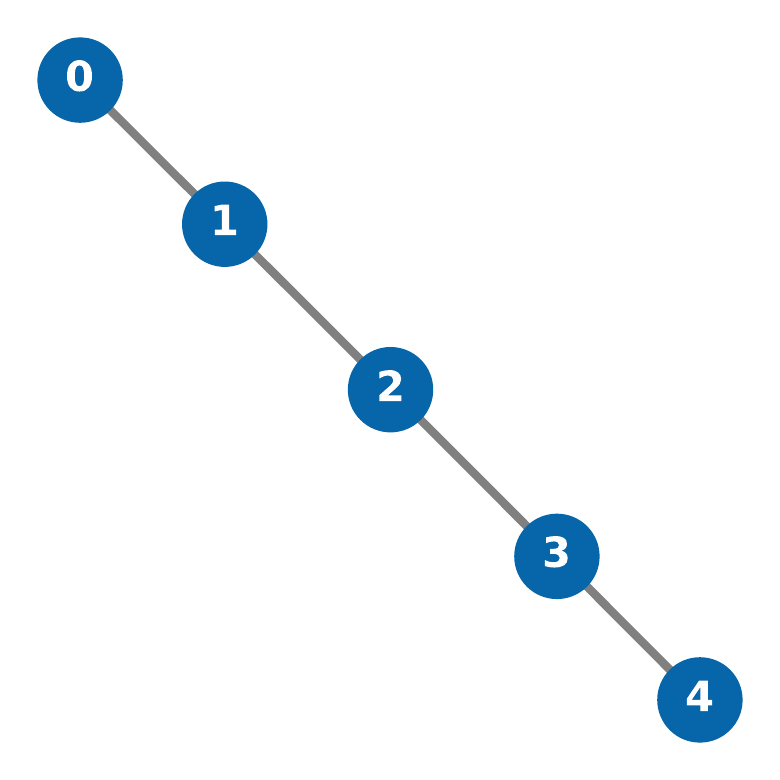}
        \caption{5L}
    \end{subfigure}
    
    \vspace{0.4cm}

    \begin{subfigure}[b]{0.32\columnwidth}
        \centering
        \includegraphics[width=\textwidth]{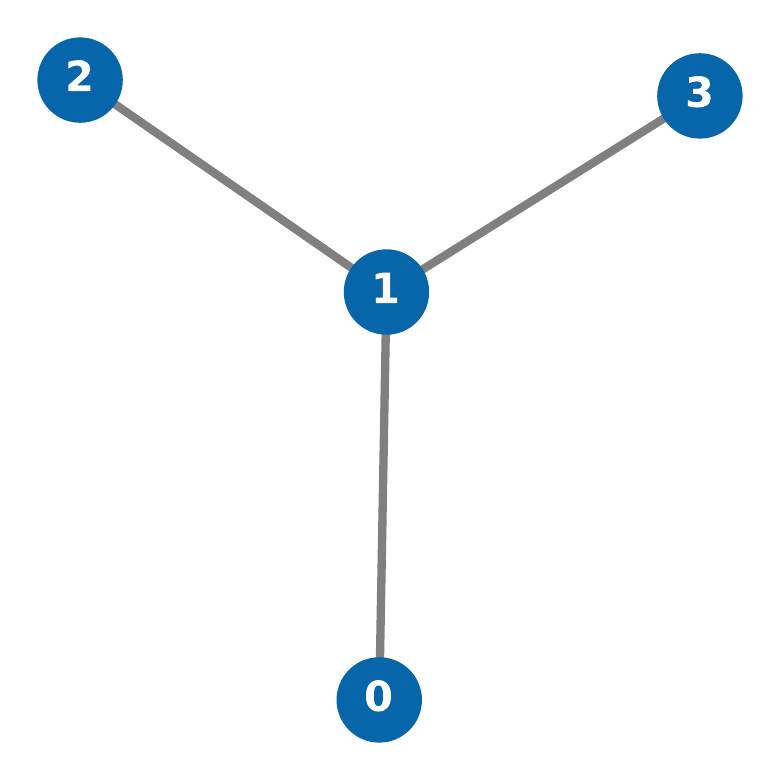}
        \caption{4T}
    \end{subfigure}\hfill
    \begin{subfigure}[b]{0.32\columnwidth}
        \centering
        \includegraphics[width=\textwidth]{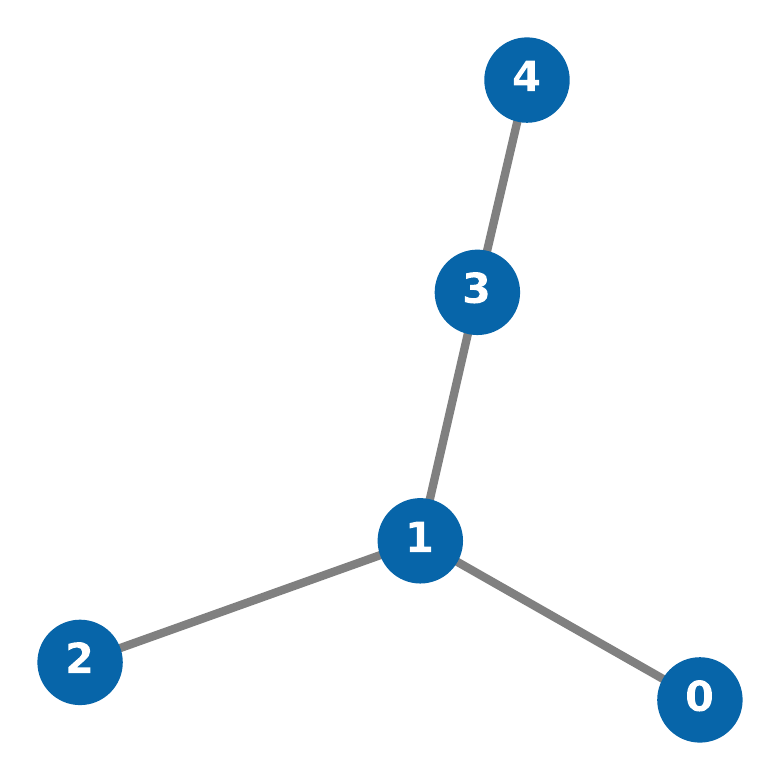}
        \caption{5T}
    \end{subfigure}\hfill
    \begin{subfigure}[b]{0.32\columnwidth}
        \centering
        \includegraphics[width=\textwidth]{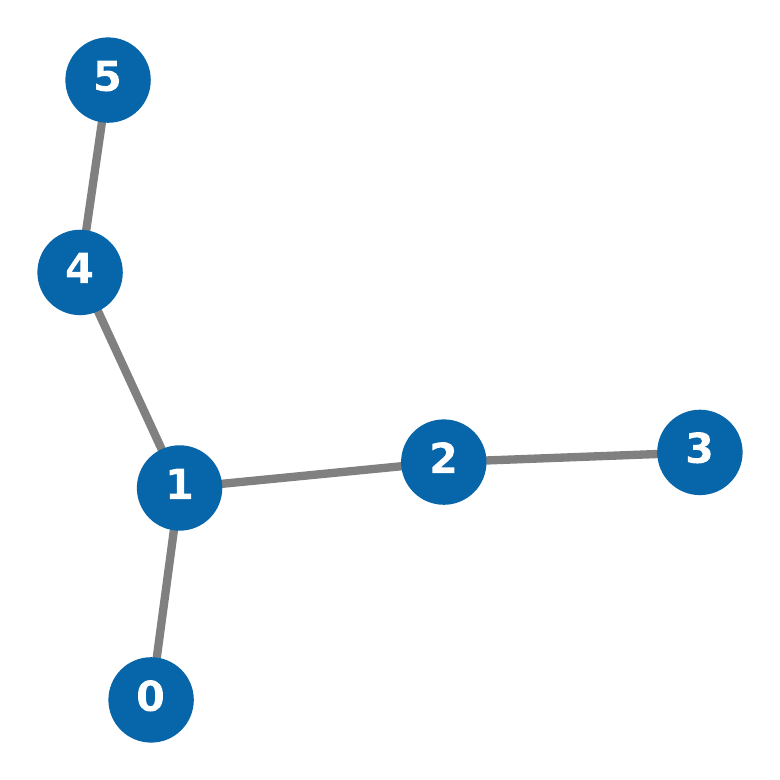}
        \caption{6T}
    \end{subfigure}
    
    \caption{Hardware connectivity maps utilized for benchmarking: Line graphs (a-c) and T-shape graphs (d-f). Vertices represent physical qubits and edges represent available coupling map connections.}
    \label{fig:topologies_2rows}
\end{figure}

Following the procedure in \cite{KremerIBM}, we select $6$ connectivity maps from $3$ to $6$ qubits (see Fig. \ref{fig:topologies_2rows} for additional information).
As before, we provide total gate count, and CNOT count. As a comparison, we evaluate  the RL-based Clifford transpiler by \texttt{qiskit}, available at \cite{qiskit_ai_transpiler_cliffords}. Both methods are sampled once and $10$ times, to take into account the stochasticity or RL-based approaches. Results are provided in Table \ref{tab:LT_topology_combined}. We observe that in general, our model provides consistently a lower total gate counts than the other approach. However, it also obtains a slightly higher number of CNOTs, suggesting that a reward function tailored on the number of two-qubit gates instead of total gates could be beneficial in this specific setting.

\subsection{Full Logical synthesis pipeline}
\label{sub:exp:pipe}

Finally, we evaluate our proposed method as a component of a logical synthesis pipeline. To do so, we first generate $10$ random circuits for each number of qubits $n\in\{5, 10, 15, 20, 25\}$. Following the approach in \cite{qpresyn}, we consider the pipeline:
\begin{itemize}
    \item Starting from the representation of the initial circuit, we decompose each 2 qubit gates in a block of CNOTs and single qubit gates with the KAK decomposition \cite{cartan_sur_1926};
    \item Then, we decompose each single qubit unitary in up to three $R_z(\theta)$ rotations (with additional $X$ and $H$ gates). Each rotation is then decomposed in $\{H, S, T\}$ with the \textsc{gridsynth} algorithm \cite{ross2016optimalancillafreeClifford} with a tolerance of $\varepsilon=0.01$; 
    \item Finally, we search for blocks of adjacent Clifford circuits. For each block of at least $2$ qubits, we use AlphaClifford to synthesize an optimized version.
\end{itemize}

\begin{table*}[!htb]
    \centering
    \begin{tabular}{lcccccccc}
        \toprule
        & \multicolumn{2}{c}{Initial Compiled} & & \multicolumn{2}{c}{Qiskit Transpile} & & \multicolumn{2}{c}{\textbf{AlphaClifford}} \\
        \cmidrule(lr){2-3} \cmidrule(lr){5-6} \cmidrule(lr){8-9}
        
        $n$ & Clifford $(k)$ & T-count $(k)$ & & Clifford $(k)$ & $\%_{red}$ & & Clifford $(k)$ & $\%_{red}$ \\
        \midrule
        \multicolumn{9}{c}{\textbf{Without Pre-compilation}} \\
        \midrule
        5  & 1.24  & 0.76  & & 1.23  & 0.36 & & 1.21  & 2.44 \\
        10 & 5.17  & 3.14  & & 5.16  & 0.13 & & 5.05  & 2.39 \\
        15 & 10.27 & 6.23  & & 10.26 & 0.09 & & 10.01 & 2.53 \\
        20 & 20.26 & 12.28 & & 20.24 & 0.14 & & 19.76 & 2.50 \\
        25 & 29.59 & 17.92 & & 29.55 & 0.14 & & 28.88 & 2.43 \\
        \midrule
        \multicolumn{9}{c}{\textbf{With Pre-compilation}} \\
        \midrule
        5  & 1.09 & 0.66 & & 1.08 & 0.35 & & 1.06 & 2.75 \\
        10 & 4.90 & 2.97 & & 4.90 & 0.17 & & 4.78 & 2.46 \\
        15 & 9.98 & 6.06 & & 9.97 & 0.15 & & 9.72 & 2.57 \\
        20 & 19.77 & 11.98 & & 19.75 & 0.12 & & 19.28 & 2.50 \\
        25 & 29.18 & 17.66 & & 29.12 & 0.20 & & 28.45 & 2.49 \\
        \bottomrule
    \end{tabular}
    \caption{Clifford synthesis performance: comparison of Clifford gate counts (in thousands) and reduction percentages relative to the original synthesized circuit. Results are grouped by whether pre-synthesis \cite{qpresyn} was applied.}
    \label{tab:logic_synthesis_restructured}
\end{table*}

In addition, we evaluate the same pipeline with the use of  \textsc{Q-PreSyn} \cite{qpresyn} to optimize the number of $T$ gates with a sequence of $2$ qubit merges. In particular, we apply a greedy optimizer to evaluate the largest sequence of gates that can be merged together, with the aim of reducing the number of resulting $T$ gates after applying the KAK decomposition followed by \textsc{gridsynth}.

In Table \ref{tab:logic_synthesis_restructured} we present the Clifford gate reduction, with the default \texttt{qiskit} transpiler as a comparison. Note that in this specific setting, due to the particular structure of the Clifford circuits (that are already optimal in the number of CNOT gates), the algorithms Aaronson-Gottesman and the ones from \cite{webster} did not obtain any reduction in total number of gates.
Finally, note that the considered pipeline applied to random circuits leads to mostly 2 and 3 qubit Clifford blocks, therefore allowing for a small reduction of gates. We expect different choices of logical synthesis algorithms  and different circuit families to provide higher reductions.

\section{Conclusion}
\label{sec:conc}

In this work, we introduced AlphaClifford, a novel model-based reinforcement learning framework designed to address the complex combinatorial challenges of Clifford circuit synthesis and transpilation. By formulating the compilation process as a Monte Carlo Tree Search over the algebraic space of symplectic matrices, AlphaClifford consistently discovers highly optimized gate sequences. 

Our experimental evaluations demonstrate that AlphaClifford achieves superior or comparable results to state-of-the-art methods across a variety of settings. For unconstrained optimization, it successfully minimizes both total and two-qubit gate counts compared to standard heuristics, even when restricted to a less expressive gate set. Furthermore, we showed its adaptability to hardware-constrained transpilation and its practical utility as a post-synthesis optimizer within a logical synthesis pipeline. Notably, these results were obtained using a single base architecture and a naive reward function, underscoring the generality and robustness of our technique. 

Ultimately, these results highlight the potential of model-based RL to transcend the limitations of traditional heuristic-driven compilers, advancing the trajectory established by frameworks like AlphaTensor \cite{alphatensor_quantum} and AlphaCNOT \cite{cossio2026alphacnotlearningcnotminimization}. As quantum hardware continues to scale toward the era of quantum utility, computationally intelligent approaches like AlphaClifford will be critical for bridging the gap between high-level logical algorithms and the strict physical constraints of near-term and future fault-tolerant quantum devices. 

We highlight several promising directions for future work. First, extending the action space to include more expressive operations, such as transvections, could further enhance compilation efficiency. Second, while our generalized reward function successfully minimized total gate counts, designing tailored reward structures—such as heavily penalizing two-qubit gates to improve specific transpilation connectivity maps or incorporating error-aware routing—could yield even better hardware-specific solutions. \nt{Finally, a further direction is to investigate systematic block-selection and iterative peephole strategies for applying the models developed here to large-scale Clifford circuits.}

\section{Acknowledgements}
DLB thanks Gianluca Macrì for his support in preparing the figures and diagrams of this work. 
    JC is supported by FSE/FVG PhD Grant on ``Computer Science and Artificial Intelligence'' (CUP G23C25000620008).

    This work has been partially supported by INdAM-GNCS project \emph{Algebra lineare quantistica, state preparation e compilazione di circuiti quantistici} (CUP E53C25002010001) and by the regional project QUASAR-FVG \emph{Calcolo e simulazione quantistica: sviluppo, applicazioni e ricerca in Friuli Venezia Giulia} (CUP G23C25001510002).

\bibliographystyle{IEEEtran}
\bibliography{bibfile}
\end{document}